\documentclass[12pt]{article}
\usepackage{hyperref,epsfig,amssymb,amsmath,graphicx,color}

\begin{document}

\baselineskip=.22in
\renewcommand{\baselinestretch}{1.2}
\renewcommand{\theequation}{\thesection.\arabic{equation}}
\newcommand{\klmt}{\mbox{K\hspace{-7.6pt}KLM\hspace{-9.35pt}MT}\ }
\begin{flushright}
{\tt ULB-TH/09-33}
\end{flushright}

\vspace{1mm}

\begin{center}
{{\Large \bf Surplus Solid Angle\\[1mm]
as\\[2mm]
an Imprint of
Ho${\check {\bf r}}$ava-Lifshitz Gravity}\\[20mm]
Sung-Soo Kim\\[2mm]
{\it Physique Th\'{e}orique et Math\'{e}matique\\
Universit\'{e} Libre de Bruxelles and International Solvay Institutes,\\
ULB-C.P. 231, B-1050 Bruxelles,
Belgium}\\
{\tt sungsoo.kim@ulb.ac.be}\\[7mm]

Taekyung Kim~~and~~Yoonbai Kim\\[2mm]
{\it Department of Physics, BK21 Physics Research Division,
and Institute of Basic Science,}\\
{\it Sungkyunkwan University, Suwon 440-746, Korea}\\
{\tt pojawd,~yoonbai@skku.edu} }
\end{center}

\vspace{8mm}

\begin{abstract}
We consider the electrostatic field of a point charge coupled to
Ho${\check {\rm r}}$ava-Lifshitz gravity and find an exact solution
describing the space with a surplus (or deficit) solid angle.
Although, theoretically in general relativity, a surplus angle is hardly
to be obtained in the presence of ordinary matter with positive energy
distribution, it seems natural in Ho${\check {\rm r}}$ava-Lifshitz gravity.
We present the sudden disappearance
and reappearance of a star image as an astrophysical effect of a surplus angle.
We also consider matter configurations of all possible power law behaviors
coupled to Ho${\check {\rm r}}$ava-Lifshitz gravity and obtain a series of
exact solutions.
\end{abstract}


\newpage

\setcounter{equation}{0}
\section{Introduction}\label{sec1}

Recently, a power-counting renormalizable and unitarity-keeping,
ultraviolet (UV) complete theory of gravity was
proposed~\cite{Horava:2009uw,Horava:2008ih,Horava:2009if} to equip space
and time with an anisotropic scaling in a Lifshitz fixed
point~\cite{Lifshitz,Horava:2008jf}. In its infrared (IR) fixed point,
the general coordinate invariance is accidentally arisen and the general
relativity is assumed to be recovered.
In this Ho${\check {\rm r}}$ava-Lifshitz (HL) gravity, subsequent
studies have been done in various directions. Specifically they cover
studies of cosmology~\cite{Takahashi:2009wc,Kiritsis:2009sh},
black hole physics and their
thermodynamics~\cite{Lu:2009em,Cai:2009pe,Colgain:2009fe,Myung:2009dc},
and other related subjects~\cite{Nikolic:2009jg}.
There have also been some discussions on the field theories in the Lifshitz
fixed point and flat space-time~\cite{Horava:2008jf,Chen:2009ka}.
In order to survive as a viable quantum gravity consistent
with observation, there remain numerous issues and challenges
to be settled down in HL gravity of which the list includes smooth IR limit vs the detailed
balance condition, the Hamiltonian dynamics in the presence of the lapse
function with spatial dependence (the projectability condition),
perturbative renormalizability, strong coupling, and propagating modes
including the degree of scalar graviton,
etc.~\cite{Sotiriou:2009gy,Charmousis:2009tc,Li:2009bg,Calcagni:2009qw,Nastase:2009nk}.

An attractive subject is to find a testable evidence given as
a unique characteristic of HL gravity, distinguishable from the properties
of GR.
Although HL gravity is assumed to reproduce GR as its IR theory,
it may be intriguing to ask which effect of HL gravity in the UV regime
is most likely to be detected through astronomical or astrophysical
observation. Following the usual strategy as is often done in GR,  one may look
for a static solution of HL theory, employ it as
a gravitational background, and investigate possible effects to a
test particle such as change of positions of the stars, which is ascribed
to HL gravity, not to GR.

In this paper we consider an electrostatic field of a point charge
as matter,
and obtain spherically symmetric solutions which describe a space with either
a surplus or deficit solid angle. The result of a surplus solid angle
from an ordinary matter with positive energy density is contrasted
with a genuine feature of GR in which it can usually be materialized
by the source of negative mass or energy.
We propose the sudden
disappearance and reappearance of stars in front of an observer
as evidence of a resultant surplus solid angle of the HL gravity theory.
Because of this qualitatively
different nature between HL gravity and GR, a single observation of a
surplus (solid) angle can possibly support or rule out HL gravity.
Although we
could not determine the specific value of a surplus solid angle yet, it must be
one of strong candidates which awaits observed data as a possible
astrophysical candidate to test HL gravity.

The paper is organized as follows. In Sec. 2 we review
the spherically symmetric static vacuum solutions.  A surplus solid
angle coming from the solution of HL gravity coupled to the electrostatic
field of a point charge is discussed in Sec. 3. We study the matters
with various power-behavior long tails and obtain exact solutions
including {\it charged} black
hole solutions in Sec. 4. We then conclude with further discussion.

\setcounter{equation}{0}
\section{Spherically Symmetric Vacuum Solutions}\label{sec2}
HL gravity~\cite{Horava:2009uw}
is a newly proposed {\it
nonrelativistic} gravity based on anisotropic scaling between
time and space
\begin{align}
t\rightarrow \ell^{z} t, \qquad x^{i}\rightarrow \ell x^{i},
\label{scaling}
\end{align}
with $z = 3$ to make the theory in (1+3) dimensions power counting
renormalizable. The metric can be written in the ADM decomposition
\begin{align}
ds^{2}=-N^2dt^2+g_{ij}\left(dx^i+N^idt\right)\left(dx^j+N^jdt\right).
\label{met1}
\end{align}

The action is given by
\begin{align}
S_{{\rm HL}} =\int dt\, d^{3}x\,\sqrt{g} N({\cal L}_{{\rm IR}}+{\cal
L}_{{\rm UV}}), \label{ac1}
\end{align}
where ${\cal L}_{{\rm IR}}$ contains the quadratic or lower
derivative terms, and ${\cal L}_{{\rm UV}}$ the higher
derivative terms
\begin{align}
{\cal L}_{{\rm IR}}=&{2\over \kappa^2}(K_{ij}K^{ij}-\lambda
K^2)+\frac{\kappa^2\mu^2 \Lambda}{8(1-3\lambda)}
\left(R-3\Lambda\right),
\label{LIR}\\
{\cal L}_{{\rm UV}}=&-{\kappa^2\over 2w^4}\left(C_{ij}-\frac{\mu
w^{2}}{2}R_{ij}\right) \left(C^{ij}-\frac{\mu w^{2}}{2}R^{ij}\right)
+\frac{\kappa^2\mu^2(1-4\lambda)}{32(1-3\lambda)}R^2. \label{LUV}
\end{align}
Here the extrinsic curvature is
\begin{align}
K_{ij}\equiv
\frac{1}{2N}\left({\dot{g_{ij}}}-\nabla_iN_j-\nabla_jN_i\right),
\quad K=g^{ij}K_{ij}, \quad K^{ij}=g^{ik}g^{jl}K_{kl}
\end{align}
with ${\dot{g_{ij}}}\equiv\frac{\partial g_{ij}}{\partial t}$ and
$\nabla_i$ covariant derivative with respect to $g_{ij}$. $R$ and
$R_{ij}$ are three-dimensional scalar curvature and  Ricci tensor,
and the Cotton tensor is given by
\begin{align}
C^{ij}=
\frac{\epsilon^{ikl}}{\sqrt{g}}\nabla_k\left({R^j}_l-\frac{1}{4}R\delta^{j}_{\;l}\right).
\label{Co1}
\end{align}

The gravity action \eqref{ac1} is invariant under a restricted class
of diffeomorphism, foliation-preserving diffeomorphism. General
coordinate invariance is regarded as an accidental symmetry in the
low energy scale with the choice of $\lambda=1$.\footnote{Taking the
Einstein limit as an IR  limit  is  actually nontrivial.
Quantization procedure of consideration~\cite{Charmousis:2009tc,
Li:2009bg} occasionally restricts the usage of the space-dependent
lapse function~\eqref{met1} and the detailed balance condition. We
will also discuss in the subsequent sections that the long distance
limit of spherically symmetric solutions of HL gravity does not lead
to those of Einstein gravity~\cite{Calcagni:2009qw, Nastase:2009nk}.
However, for this classical static problem, we will keep the
detailed balance condition to make the problem tractable.}
There are seven independent terms in the action \eqref{ac1}, yet the parameters in HL gravity are five, $\kappa
,~\lambda ,~\Lambda,~w ,~\mu$ thanks to the detailed balance.
 When the IR action \eqref{LIR} is
directly compared to the (1+3)-dimensional Einstein-Hilbert action
with speed of light $c$, Newton's gravitational constant $G$, and
the effective cosmological constant $\Lambda_{{\rm E}}$, three parameters can be determined in addition
to $\lambda=1$,
\begin{align}
c=\frac{\kappa^2\mu}{4}\sqrt{\frac{\Lambda}{1-3\lambda}},\quad 16\pi
G= \frac{\kappa^2}{2c},\quad \Lambda_{{\rm E}}= \frac{3}{2}\Lambda .
\label{con1}
\end{align}
In the presence of matter fields, the matter action takes the form
\begin{equation}\label{acma}
S_{\rm M}=\int dtd^3x \sqrt{g}N~{\cal L}_{{\rm M}}(N,N_i,g_{ij}).
\end{equation}

We now restrict ourselves to the static solutions with spherical
symmetry. Let us choose a reference frame and introduce spherical
coordinates $(t,r,\theta,\phi)$ with a static
metric
\begin{equation}
ds^2=-B(r)e^{2\delta(r)}dt^2+\frac{dr^2}{B(r)}+r^2(d\theta^2+\sin^2\theta
d\varphi^2). \label{rmet}
\end{equation}
The projectability condition is not taken into account
in this work, although it is necessary especially for quantization of the HL
gravity~\cite{Horava:2009uw}. In order to fulfill it automatically it is
convenient to use Painlev{\' e}-Gullstrand type
coordinates~\cite{Sotiriou:2009gy}, which will be presented
elsewhere~\cite{KimKim}.

Since all the components of Cotton tensor vanish, $C^{ij}=0$ under
this metric, we can write the action \eqref{ac1} in
a simple form
\begin{align}
S_{{\rm HL}} =&  \frac{\pi\kappa^2\mu^2}{2(3\lambda-1)} \int dt\int dr \,
e^{\delta}\times
\nonumber\\
&\Bigg\{ (1-3\lambda) \Bigg[{\tilde B}'^2 +2\Big(\frac{\tilde B}{r}
+\frac{{\tilde B}'}{2}\Big)^2 \Bigg] -(1-4\lambda)\Big(\frac{{\tilde
B}}{r}+{\tilde B}'\Big)^2 +2\Lambda r\Big(\frac{{\tilde
B}}{r}+{\tilde B}'\Big)+3\Lambda^2r^2 \Bigg\}, \label{rac}
\end{align}
where we introduced ${\tilde B}=B-1$. The equations of motion are
then readily obtained by varying the action with respect to the
metric functions
\begin{align}
&~\Bigg[(\lambda-1){\tilde B}'-\frac{2\lambda}{r}{\tilde B}
-2\Lambda r\Bigg]\delta'+ (\lambda-1)\tilde B'' -
\frac{2(\lambda-1)}{r^{2}}\tilde B
=\frac{8(1-3\lambda)r^2}{\kappa^2\mu^2}\frac{\partial {\cal L}_{\rm
M}}{\partial B},
\label{Beq}\\
&~ (1-3\lambda) \Bigg[ {\tilde B}'^2+2\Big( \frac{\tilde B}{r}
+\frac{\tilde B'}{2} \Big)^2 \Bigg] -(1-4\lambda)\Big(\frac{\tilde
B}{r}+{\tilde B}'\Big)^2 +2\Lambda r \Big(\frac{\tilde B}{r}+{\tilde
B}'\Big)+3\Lambda^2 r^{2}
\nonumber\\
&\hspace{85mm}=\frac{8(1-3\lambda)r^{2}}{\kappa^2\mu^2}\left({\cal
L}_{\rm M} +\frac{\partial {\cal L}_{\rm M}}{\partial
\delta}\right), \label{deleq}
\end{align}
where we included the matter action for future purpose
\begin{equation}
S_{\rm M}= 4\pi \int_{-\infty}^{\infty}dt\int_{0}^{\infty} dr r^2
e^{\delta} {\cal L}_{\rm M}(B,\delta) . \label{acma1}
\end{equation}

It was claimed 
that $\lambda$ flows to unity in the low
energy regime and HL gravity reduces to the Einstein theory with
a negative cosmological constant as long as the higher order
curvature terms, ${\cal O}(R^{2}, R_{ij}^{2}, ...)$, are neglected.
In the Eqs. \eqref{Beq} and \eqref{deleq}, these correspond to
insertion of $\lambda=1$ and neglect of the terms quadratic in
the metric functions, ${\tilde B}^{2}$ and ${\tilde B}\delta$. Then,
without matter ${\cal L}_{{\rm M}}=0$, the equations reduce to the
Einstein equations and then yield anti-de Sitter (AdS) Schwarzchild solution with integration constants
$\delta_0$ and $M$,
\begin{align}
r\frac{d\delta}{dr}=0, \qquad\qquad\qquad~~\longrightarrow&\quad
\delta(r)=\delta_0=0,
\label{Beq2}\\
\frac{d}{dr}\left(rB\right)=1-\frac{3\Lambda}{2}r^2,
\quad~\longrightarrow&\quad
B(r)=-\frac{\Lambda}{2}r^2+1-\frac{M}{r}. \label{deleq2}
\end{align}

The known exact solutions to  \eqref{Beq} and \eqref{deleq} are as follows \cite{Lu:2009em, Myung:2009dc}.
The generic solution is $B(r)=1-\Lambda r^{2}$, which is independent of $\lambda$.
For $\lambda\ge 1/3$ and $\lambda\neq1$, there exists another solution
\begin{align}
B(r)=1-\Lambda r^{2}+ B_{0} r^{\frac{2\lambda \pm\sqrt{2(3\lambda-1)}}{\lambda-1}},\qquad
\delta(r)= \frac{1+3\lambda\mp2\sqrt{(3\lambda-1)}}{1-\lambda}\ln(r/r_{0}),
\end{align}
and, for $\lambda=1$,
\begin{align}
B(r)=1-\Lambda r^{2}\pm\sqrt{c_{1}r}\ , \qquad \delta(r)=\delta_{0}=0, \label{lambda1}
\end{align}
where $B_{0}$ and $c_{1}$ are integration constants. There are also some special solutions depending on the value of $\lambda$, see the appendix.

All the known exact solutions consistently show that they
cannot reproduce the AdS Schwarzschild solution
\eqref{Beq2} and \eqref{deleq2} in the long distance limit. After
substituting $\lambda=1$, the linear Einstein equations
\eqref{Beq2} and \eqref{deleq2} are obtained from the nonlinear
equations \eqref{Beq} and \eqref{deleq} without matter, by utilizing
$|\Lambda r|\gg |{\tilde B}^{'}|\sim |{\tilde B}|/r$. This
assumption is not consistent with the leading long distance behavior
of $B(r)$ in \eqref{ccso}. Therefore, the nonlinear terms cannot be
neglected even in a long distance limit, and it may support possible
mismatch between the IR limit of HL gravity and the Einstein
theory in the context of classical solutions.

The solution for $\lambda=1$ is of interest. For the lower minus
sign in \eqref{lambda1}, it is easy to see that the number of horizons
varies from zero to two depending on the relation between $c_{1}$
and $\Lambda$.
To see physical singularity in the spatial sector, we examine the
three-dimensional analogue of the Kretschmann invariant
\begin{align}
R^{ijkl}R_{ijkl}=4 R_{ij}R^{ij}-R^2
=\frac{2}{r^2}\left[B'^{\,2}+2\left(\frac{B-1}{r}\right)^2\right],
\label{Kre}
\end{align}
where we used a three-dimensional identity,
$R_{ijkl}=g_{ik}R_{jl}-g_{il}R_{jk}-g_{jk}R_{il}+g_{jl}R_{ik}-\frac{1}{2}
(g_{ik}g_{jl}-g_{il}g_{jk})R$. The $\lambda=1$ solution \eqref{lambda1} then
has singularity,
$R^{ijkl}R_{ijkl}=\frac{9c_1}{2r^{3}} \mp \frac{12 \Lambda
\sqrt{c_1}}{r^{3/2}} +12 \Lambda^{2}$,
unlike the first solution independent of $\lambda$. For sufficiently large $r$, we have a
constant piece of the curvature, whose value is
different from the constant curvature of the AdS Schwarzschild
solution.

We note that the absence of singularity does not guarantee that
other observers do not see singularity at all. By the same token,
although a singularity is obtained from \eqref{Kre}, it is subtle
whether the singularity still appears as a singularity to other
observers. Since time reversal, time translation, parity, and
spacial rotation are symmetries of the system, observers who are
connected by these symmetries surely see the singularity. However,
if we take into account that boost is not a symmetry, then it is not
clear how the singularity is observed to those who are in a relative
motion.

We close this section with a remark. In this nonrelativistic HL gravity with speed of light
larger than that in general relativity, horizons and curvature singularities can be dependent
upon the choice of reference frames, and therefore the meaning of
these is different from those in general relativity with general coordinate invariance.

\setcounter{equation}{0}
\section{Electrostatic Field and Surplus (Deficit) Solid Angle}\label{sec3}

The action of U(1) gauge theory in the Lifshitz fixed point with $z=2$ is
constructed in flat space and time by using detailed
balance~\cite{Horava:2008jf},
\begin{align}
S_{{\rm U(1)UV}}=\int dtd^{3}x\left[\frac{1}{2}E_{i}^{2}
-\frac{1}{8g^{2}}(\partial_{i}F_{ik})(\partial_{j}F_{jk})\right],
\label{hema}
\end{align}
where $E_{i}=F_{i0}$ is electric field and
$F_{ij}=\partial_{i}A_{j}-\partial_{j}A_{i}$.
This vector field theory possesses neither (Galilean) boost symmetry
nor electromagnetic duality even in the absence of external sources.
In IR regime, both Lorentz symmetry and duality are accidentally arisen, and
the action is
\begin{align}
S_{{\rm U(1)IR}}=-\frac{1}{4}\int dtd^{3}xF_{\mu\nu}F^{\mu\nu},
\label{IRM}
\end{align}
where $F_{\mu\nu}=\nabla_{\mu}A_{\nu}-\nabla_{\nu}A_{\mu}$.
Along the flow line from the UV fixed point to the IR fixed point,
generic action is assumed to take the following form,
\begin{align}
S_{{\rm U(1)}}=\int dtd^{3}x\left[\frac{1}{2}E_{i}^{2}
-F_{{\rm U(1)}}(F_{ij},\partial_{i}F_{jk})\right],
\label{U1}
\end{align}
where the {\it potential} can contain all possible gauge-invariant
combinations of $F_{ij}$ and $\partial_{i}F_{jk}$ as long as it
satisfies $F_{{\rm
U(1)}}=-\frac{1}{8g^{2}}(\partial_{i}F_{ik})(\partial_{j}F_{jk})$ at
the Lifshitz fixed point and $F_{{\rm U(1)}}\rightarrow
-\frac{1}{4}F_{ij}^{2}$ in the IR limit. If the theory in the UV
limit \eqref{hema} flows to the IR limit \eqref{IRM} keeping the
detailed balance, the form of the action in the intermediate scale
\eqref{U1} is restricted. Although we discussed a specific
anisotropic scaling of $z=2$, the above discussion can
straightforwardly be generalized to the cases of arbitrary $z$ (see
the $z=3$ case given in Ref.~\cite{Kiritsis:2009sh}).

Unlike the magnetic potential terms which change and become
complicated along the flow line, the kinetic term of electric field
is unaltered. For this reason, we consider only the electrostatic
field due to a point object of electric charge $q_{{\rm e}}$ at the
origin, and the magnetic potential will be discussed in the
following section.

Using the same static metric \eqref{rmet}, we find that the
equations of motion are given in terms of nonvanishing components
\begin{align}
\frac{\partial {\cal L}_{\rm M}}{\partial B}=0 , \qquad {\cal
L}_{\rm M}+\frac{\partial {\cal L}_{\rm M}}{\partial
\delta}=-\frac{1}{2}e^{-2\delta} F_{r0}^{2} =
-\frac{1}{32\pi^2}\frac{q_{{\rm e}}^2}{r^{4}}, \label{mat5}
\end{align}
where we used
\begin{align}
E_{r}=F_{r0}=\frac{e^{\delta}q_{{\rm e}}}{4\pi r^{2}} . \label{emfd}
\end{align}
In the limit of Einstein theory, the Eqs. \eqref{Beq} and \eqref{deleq}
reduce to
\begin{align}
r\frac{d\delta}{dr}=0, \qquad&\longrightarrow\quad
\delta(r)=\delta_0=0,
\label{Beq22}\\
\frac{d}{dr}\left(rB\right)=1-\frac{3\Lambda}{2}r^2
-\frac{1-3\lambda}{8\pi^2\kappa^2\mu^2 \Lambda}\frac{q_{\rm
e}^2}{r^2}, \quad&\longrightarrow\quad
B(r)=-\frac{M}{r}+1-\frac{\Lambda}{2}r^2+\frac{1-3\lambda}{8\pi^{2}
\kappa^2\mu^2 \Lambda}\frac{q_{\rm e}^2}{r^2}. \label{deleq22}
\end{align}
It follows from \eqref{con1} that  $\Lambda<0$ leads to
$\lambda>1/3$ which gives the following asymptotic behaviors
\begin{align*}
\lim_{r\rightarrow \infty}B(r)\sim \frac{|\Lambda|}{2} r^{2}
\rightarrow \infty\quad {\rm and}\quad \lim_{r\rightarrow 0}B(r)\sim
\left|\frac{1-3\lambda}{8\pi^{2} \kappa^2\mu^2 \Lambda}\right|
\frac{q_{\rm e}^2}{r^2} \rightarrow \infty ,
\end{align*}
and thus the maximum number of horizon is four. Let us restrict ourselves to
the limit $\Lambda\rightarrow0$ keeping $M>0$ and
$\kappa^{2}\mu^{2}\Lambda$ finite. When $\lambda>1/3$ and
$\Lambda\rightarrow0^{-}$, one then finds that horizons are given as
\begin{align}
\begin{cases}
\displaystyle{ r_{{\rm
H}}=\frac{1}{2}\left[M\pm\sqrt{M^{2}-\frac{(1-3\lambda)q_{{\rm
e}}^{2}}{ 2\pi^{2}\kappa^{2}\mu^{2}\Lambda}}\,\right]},&
\quad\mbox{when}\quad \displaystyle{ M^{2}>\frac{(1-3\lambda)q_{{\rm
e}}^{2}}{2\pi^{2}\kappa^{2}\mu^{2}\Lambda}
}\\
\displaystyle{ r_{{\rm H}}=\frac{M}{2},} & \quad\mbox{when}\quad
\displaystyle{ M^{2}=\frac{(1-3\lambda)q_{{\rm
e}}^{2}}{2\pi^{2}\kappa^{2}\mu^{2}\Lambda} }
\\
\mbox{no~horizon}, & \quad\mbox{when}\quad \displaystyle{
M^{2}<\frac{(1-3\lambda)q_{{\rm
e}}^{2}}{2\pi^{2}\kappa^{2}\mu^{2}\Lambda} }
\end{cases}
\, .
\end{align}
The first case reproduces the AdS Reissner-Nordstr\"{o}m black hole
and the second case is its extremal limit. For $\lambda=1/3$, the
coefficient of $1/r^{2}$ term in \eqref{deleq22} vanishes and thus
leads to the limit of Schwarzschild black hole. When $\lambda<1/3$
and $\Lambda\rightarrow0^{+}$, only one horizon is formed at
\begin{align}
r_{{\rm H}}=\frac{1}{2}\left[M+\sqrt{M^{2}-\frac{(1-3\lambda)q_{{\rm
e}}^{2}}{ 2\pi^{2}\kappa^{2}\mu^{2}\Lambda}}\,\right] .
\end{align}

By solving the full Eqs. \eqref{Beq} and \eqref{deleq} including the
matter \eqref{mat5}, we obtain an exact solution for $1/3\le\lambda<1/2$,
\begin{align}
B(r)=1\pm \Delta_{q_{\rm e},\lambda}-\Lambda r^{2}, \qquad
\delta(r)=\frac{1-\lambda}{\lambda}\ln (r/r_{0}), \label{ch1}
\end{align}
where
\begin{align}
\Delta_{q_{\rm e},\lambda}=\sqrt{\frac{1-3\lambda}{2\lambda-1}}\,
\frac{|q_{{\rm e}}|}{2\pi\kappa\mu}. \label{De}
\end{align}
In the absence of electric charge $q_{{\rm e}}=0$, the solution
reproduces a vacuum solution \eqref{lambda1} as expected.
Unlike the vacuum solutions,
there is no special solution for $\lambda=1$.
\footnote{ In the
presence of  electromagnetic field of electric charge $q$ and
magnetic monopole charge $p$, another solution for $\lambda=1$ was
claimed to exist in Ref.~\cite{Colgain:2009fe},
\begin{align}B(r)=1-\Lambda r^2\pm\sqrt{\Delta_{q_{\rm e},1}^2+c_1
r}\,,\qquad \delta(r)=\delta_0=0 \label{ch4}
\end{align}
with $\Delta_{q_{\rm e},1}=\sqrt{\frac{8(q^2+p^2)}{(\kappa\mu
g)^2}}$ and an integration constant $c_1$. The difference is that
the authors of ~\cite{Colgain:2009fe} used analytic continuations to make the cosmological
constant positive whereas we have not. This results in overall sign difference in the gravity sector.
On the other hand, the matter sector in \cite{Colgain:2009fe} needs some modifications.
To obtain solutions, one should solve equations of motion for metric functions and a vector field simultaneously.
Hence,
the matter part of the Lagrangian density in
(14) of Ref.~\cite{Colgain:2009fe} should be corrected as
\begin{align}
\frac{N}{r^2 \sqrt{f}}(q^2+p^2)~\longrightarrow~&\frac{N
r^2}{\sqrt{f}}\left(\frac{f}{N^2}F_{r0}^2-\frac{F_{\theta\phi}^2}{r^4\sin^2\theta}\right)\label{CY1}\\
&=\frac{N}{r^2\sqrt{f}}\left(q^2-p^2\right),\label{CY2}
\end{align}
and one then varies the action with respect to $N$ and $f$ in \eqref{CY1} instead of those
in \eqref{CY2}.
}

Notice that the electrostatic field couples to HL gravity gives
a constant contribution $\pm\Delta_{q_{\rm e},\lambda}$ in
\eqref{ch1},
 unlike the $1/r^{2}$ contribution of \eqref{deleq22} in Einstein gravity.
This may lead to a surprising consequence that is not expected in
Einstein gravity. Indeed, via the following rescaling
\begin{align}
dt\rightarrow \left(1\pm \Delta_{q_{{\rm
e}},\lambda}\right)^{\frac{1}{2\lambda}}dt, \qquad dr\rightarrow
\sqrt{1\pm \Delta_{q_{{\rm e}},\lambda}}\, dr ,
\end{align}
one can rewrite the metric with \eqref{ch1} as
\begin{align}
ds^{2}=-(1-\Lambda r^{2})
(r/r_{0})^{\frac{2(1-\lambda)}{\lambda}}dt^{2}
+\frac{dr^{2}}{1-\Lambda r^{2}}+r^{2}\left(1\pm \Delta_{q_{{\rm
e}},\lambda}\right) (d\theta^2+\sin^2\theta d\varphi^2),\label{met2}
\end{align}
which describes a space with a {\it surplus (or deficit) solid
angle}.

When $0<\Delta_{q_{{\rm e}},\lambda}<1$, the lower minus sign gives
an AdS space with a deficit solid angle $4\pi\Delta_{q_{{\rm
e}},\lambda}$~\cite{Barriola:1989hx,Kim:1996pa}. The upper plus
sign, on the other hand, yields an AdS space with a surplus solid
angle;
in the limit of vanishing $\Lambda$, the area of a sphere of radius
$r$ is not $4\pi r^{2}$, but larger [see Fig.~\ref{fig1}-(a)].

\subsection{Effects of Surplus Solid Angles}
\begin{figure}[h]\centering
\scalebox{1.1}[1.1]{
\includegraphics[width=60mm]{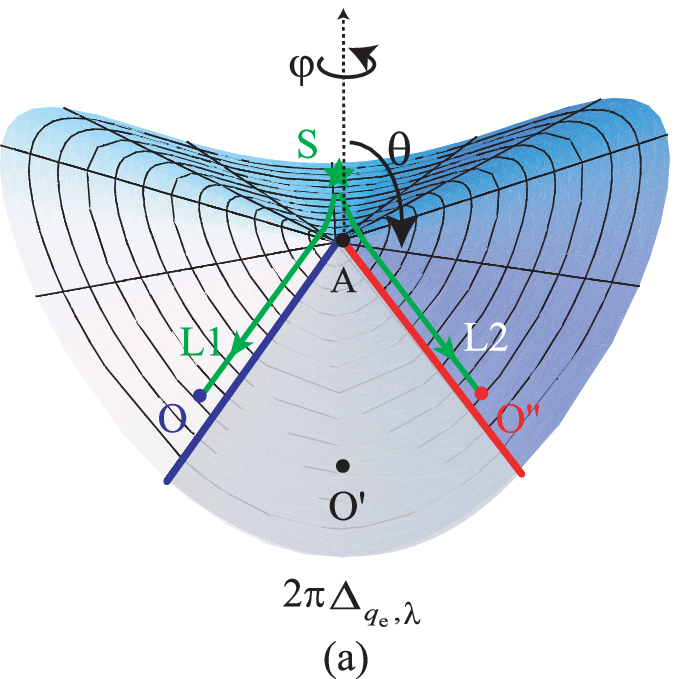}\hspace{10mm}\includegraphics[width=70mm]{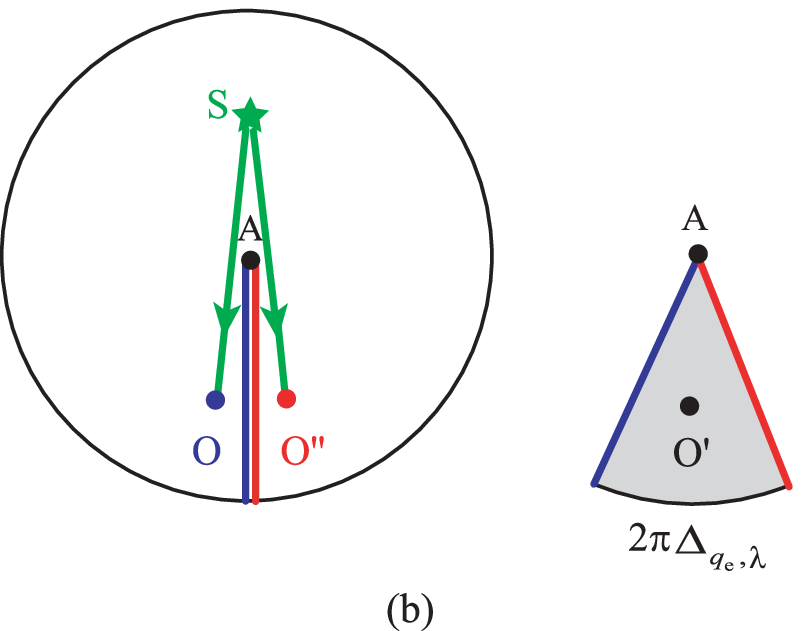}}
\\
\caption{\small The sudden disappearance of a star image due to a surplus solid
angle:
(a) Folded geometry of an {\it excess} cone representing
$\theta=\frac{\pi}{2}$ of three-dimensional flat space with surplus
angle $4\pi\Delta_{q_{{\rm e}},\lambda}=\frac{\pi}{2}$,
(b) Its development figures where the surplus solid angle region is given by
grey area.}
\label{fig1}
\end{figure}

It is instructive to consider its development figures, to understand geometry with a surplus angle. For convenience, let us consider
flat space and time limit, $\Lambda\rightarrow 0$, and set
$\theta=\pi/2$. In Fig.~\ref{fig1}-(b), the left figure is
$\mathbb{R}^{2}$ (white color) plane with a cut of a half straight
line connected to a point charge at the apex A, and a pie-slice
(grey color) on the right is a surplus piece. To make a surplus
angle like Fig.~\ref{fig1}-(a), one inserts a pie-slice into the cut
and glues the lines of the same color together.

We then investigate possible effects of the surplus solid angle
based on the development figures. We introduce a singularity at the
apex A, a star S, and three observers scattered at different points
O, ${\rm O}^{'}$, and ${\rm O}^{''}$ but connected by spatial
rotations along the angle $\varphi$ as shown in Fig.~\ref{fig1}.
Imagine that two light rays in green, L1 and L2, are emitted from a
star and reach to observers. The L1 passes through the left side of
the apex A and arrives at the observer O, the L2 reaches to the
observer O$''$ after passing the right side of the apex A. The
observers O and ${\rm O}^{''}$ look very close to each other in the
development figure, all of the observers at O,  ${\rm O}^{'}$, ${\rm
O}^{''}$, are actually separated apart shown in Fig.~\ref{fig1}-(a).
 Suppose that the singularity source at the apex A is opaque as charged stars usually do,
 then it is very likely that the third observer ${\rm O}^{'}$ in the grey colored
development figure cannot observe any light from the star  S
\footnote{A charged object can be bright, but the color of the
object at A can be assumed to be different from that of the star S
as usual. An important point is that this grey colored region due to the
surplus angle, where the star image disappears, does not depend on
the size of the opaque object at A.}. It means that any observer at
the grey colored region cannot see the image of the star at S.

We may also see the effects of a surplus angle in the following way.
Imagine this time that a star is slowly moving from O through
${\rm O}^{'}$ to ${\rm O}^{''}$, and there is a static observer at S
tracking down the trajectory of the star.  As the star once passes
through the red line, the star image may disappear for a while to an
observer at S and reappear at a distant point.

If we imagine a drastic case that a charged astronomical object is
suddenly created (or annihilated) and such event happens at the apex
A, then all the stars in the entire grey region disappear (or
appear) in front of the observer at S in the twinkling of an eye.
This sudden disappearance of a star image can be a testable
astronomical or astrophysical effect of the surplus solid angle.

In the aforementioned argument we used only the straight propagation
property of light in flat space and the speed of light does not
affect the obtained result.
To confirm, we comment on how speed of light changes by checking lightlike
geodesics.
The Lagrangian $L$ for the trajectory of light associated with the metric
\eqref{met2} reads
\begin{align}
L=&-(1-\Lambda r^{2}) (r/r_{0})^{\frac{2(1-\lambda)}{\lambda}}{\dot
t}^{2} +\frac{{\dot r}^{2}}{1-\Lambda r^{2}}+r^{2}\left(1\pm
\Delta_{q_{{\rm e}},\lambda}\right) ({\dot \theta}^2+\sin^2\theta
{\dot \varphi}^2),
\end{align}
where the dot is the derivative with respect to affine parameter.
This yields two constants of motion, $E=2(1-\Lambda
r^2)(r/r_{0})^{\frac{2(1-\lambda)}{\lambda}} {\dot t}>0$ and
$\ell =2r^2(1\pm \Delta_{q_{{\rm e}},\lambda})\sin^2\theta{\dot \varphi}$.
Since we are interested in the light radially emitted from a star S,
there no angular momentum, $\ell=0$,
and then the azimuth angle is also a constant of motion. Choosing
$\theta=\pi/2$ and $\dot\theta=0$, we find that for the light with
vanishing Lagrangian ($L=0$),
\begin{align}
{\dot r}=\pm\frac{E}{2(r/r_0)^{\frac{1-\lambda}{\lambda}}},
\qquad(1/3\leq\lambda<1/2),
\end{align}
which shows that  the speed of light is ${\dot r}_0=E/2$ by placing S
at $r_{0}$, and more importantly that it diverges as light approaches
the apex A and decreases  when traveling away from the apex.

When $\Delta_{q_{{\rm e}},\lambda}\ge 1$ with
the lower minus sign, the metric function \eqref{ch1} is rewritten as
\begin{align}
B(r)=-\Lambda r^{2}-M~~\mbox{with}~~M=\Delta_{q_{{\rm
e}},\lambda}-1\ge 0.
\label{ch3}
\end{align}
A horizon is at $r_{{\rm H}}=\sqrt{M/(-\Lambda)}\,$, and the
metric takes the same form as that of (1+2)-dimensional BTZ black hole
of mass $M$ which is now proportional to the electric charge $|q_{{\rm e}}|$.
Thus, both a resultant surplus (or deficit) solid angle and mass $M$ are a quantity
that is proportional to
the absolute value of electric charge $|q_{{\rm e}}|$. It means that
the origin of those observable quantities in a given frame is not
an arbitrary integration constant, but an externally given charge.
In Einstein gravity, mass and (electric) charge are
independently observed black hole charges. In a given frame these two
quantities which appear in the
solution~\eqref{ch3} of HL gravity for various $\lambda$,
 hardly seem to be distinguished in these classical configurations.

In Einstein gravity with the linear equation of $B(r)$
\eqref{deleq22}, the deficit solid angle is formed by a global
monopole of which energy density is proportional to $1/r^{2}$ for
large $r$~\cite{Barriola:1989hx}. In order to obtain a surplus solid
angle in the context of Einstein theory, we need exotic negative
energy density everywhere, {\it e.g,} a $\delta$-function like
distribution of antimatter with negative mass in (1+2) dimensions,
and linearly divergent negative energy distribution (like a global
monopole) in (1+3) dimensions. But these violate the positive energy
theorem. In HL gravity, however, both the surplus and deficit solid
angles are naturally obtained from ordinary electrostatic field of
which energy density is always positive. This implies that
astrophysical observation of a sudden jump of a star image due to a
surplus solid angle may become an imprint of HL gravity.

Although the realistic value of surplus (or deficit) solid angle tempts
us as
\begin{align}
4\pi\Delta_{q_{{\rm e}},\lambda}=\sqrt{\frac{\Lambda_{{\rm
E}}}{3(2\lambda-1)} \frac{16\pi G}{c}}\, |q_{{\rm e}}| \label{vDe}
\end{align}
by using \eqref{con1}, the obtained solution \eqref{ch1} is not
connected to the AdS charged black hole solution of Einstein theory
\eqref{Beq22}--\eqref{deleq22} in the long distance limit, and thus
a direct usage of the values of $c$, $G$, $\Lambda_{{\rm E}}$ in the
present universe seems not to be natural.

It is also worth noting that one can easily find a singularity at the origin,
inversely proportional
to the fourth order of the radial coordinate as expected,
\begin{align}
R^{ijkl}R_{ijkl}= 4\left(\frac{\Delta_{q_{{\rm
e},\lambda}}}{r^{2}}\pm\Lambda\right)^{2} +8\Lambda^{2},
\end{align}
where \eqref{ch1} is used.
This singularity is due to the divergent electrostatic field
\eqref{emfd} from a point electric charge, not directly from the
$\delta$-function-like singularity of the point charged particle
itself. In nonrelativistic HL gravity an observer seems possibly to
hit the singularity.

\setcounter{equation}{0}
\section{Monopoles of Anisotropic Scalings}\label{sec4}

In the previous section we dealt with the kinetic term composed of
electric field of net charge. In this section we consider static
gravitating monopoles which are supported from the potential. The
potential terms in HL gravity allow spacial derivatives whose form
depends on the anisotropic scaling and changes along the flow from the UV fixed point to the IR limit.

Before we examine the derivatives and potentials for monopoles, we
first explore possible profile of the monopole configurations. For
both global monopoles of O(3) linear sigma model and magnetic
monopoles of U(1) gauge theory in the HL type field theories, the
long distance behavior of the Lagrangian density must exhibit
power-law  behavior as ${\cal L}_{{\rm M}}\sim 1/r^{n}$ irrespective
of the value of $z$. To be specific, we consider
\begin{align}
\frac{\partial {\cal L}_{\rm M}}{\partial B}\approx 0,\qquad {\cal
L}_{\rm M}+\frac{\partial {\cal L}_{\rm M}}{\partial \delta} \approx
-\frac{\alpha}{r^{n}}, \qquad (n=0,1,2,...), \label{M4}
\end{align}
where $\alpha$ is a constant to be determined by the explicit form
of matter Lagrangian and the monopole configurations of interest. In
particular,  we are interested in non-negative integer $n$,
otherwise the matter distribution diverges at spatial infinity, and
positive $\alpha$ to ensure positivity of the matter energy
density.\footnote{If we use
$T_{\mu\nu}=-\frac{2}{N\sqrt{-g}}\frac{\delta S_{\rm M}}{\delta
g^{\mu\nu}}$ as the energy-momentum tensor of matter fields, we
obtain the energy density for the spherically symmetric matter
distribution and the metric \eqref{rmet}
\begin{align*}
-T^{0}_{\,\,0}=-2\pi\left({\cal L}_{\rm M}+\frac{\partial {\cal
L}_{\rm M} }{\partial \delta}\right) \qquad {\rm with} \quad
\frac{\partial {\cal L}_{\rm M} }{\partial B}=0.
\end{align*}
Non-negativity of the energy density distribution from the matter
fields of our interest requires non-negative $\alpha$ in \eqref{M4}.
We may not accept positive $\alpha$ to get a physical solution
except for the case of a negative cosmological constant ($n=0$),
otherwise it leads to negative energy density in the entire space
given in $(r,\theta,\phi)$ coordinates. Since the time-derivative
terms in the action are quadratic, the aforementioned
$-T^{0}_{\,\,0}$ is the same as canonical Hamiltonian density by
Legendre transform. } It follows from \eqref{Beq} and \eqref{deleq}
that exact solutions are given as follows. When $n\ne 6$ and
$\frac{1}{3}\le \lambda<\frac{1}{3}+\frac{2(n-6)^{2}}{3n^{2}}$, we
find
\begin{align}
B(r)=1-\Lambda
r^2\pm\Delta_{\alpha,\lambda}\,\,r^{\frac{4-n}{2}},\qquad
\delta(r)=\frac{n(n-6)(\lambda-1)}{2[4+n(\lambda-1)]}\,\ln
(r/r_{0}),\label{gs1}
\end{align}
where
\begin{align}
\Delta_{\alpha,\lambda}=\sqrt{\frac{8(1-3\lambda)\alpha}{
\left\{n\left[1+\frac{n(\lambda-1)}{8}\right]-3\right\}\kappa^{2}\mu^{2}}}
\ .
\label{Dn}
\end{align}
When $n=6$, solution exists only when $\lambda=1/3$. It follows from
\eqref{Beq} and \eqref{deleq} that $\lambda=1/3$ yields vanishing
matter contributions and thus the solution is the same as a vacuum
solution with $\lambda=1/3$. We also find special solutions for
$\lambda=1$,
\begin{align}
&B(r)=1-\Lambda r^2\pm\sqrt{\Delta_{\alpha,1}^2r^{4-n}+c_1
r}\,,\qquad \delta(r)=\delta_0=0, \quad (n\neq3),
\label{gs2}\\
&B(r)=1-\Lambda r^2\pm\sqrt{\frac{16\alpha}{\kappa^2\mu^2} \,\, r
\ln r+c_1 r}\,,~\quad\delta(r)=\delta_0=0,\quad (n=3) \label{gs3}
\end{align}
with an integration constant $c_{1}$. Note that the $n=4$ solutions
in \eqref{gs1} and \eqref{gs3} correspond respectively to
\eqref{ch1} and \eqref{ch4} of the electrostatic field discussed in
the previous section. Since the Lagrangian density \eqref{M4} is
singular only at the origin for any positive $n$, the spatial
Kretschmann invariant has singularity at $r=0$ for all the solutions
of $n>0$,
\begin{align}
R_{ijkl}R^{ijkl}=
\begin{cases}
\displaystyle{
4\left[\left(\frac{n^2}{8}-n+3\right)\Delta_{\alpha,\lambda}^2r^{-n}\pm
(n-6)\Lambda\Delta_{\alpha,\lambda}r^{n/2}+3\Lambda^2\right] } &
\mbox{for}~\eqref{gs1}
\\
\displaystyle{
8\left[\frac{c_1+(4-n)\Delta_{\alpha,1}^2r^{3-n}}
{4r\sqrt{c_1r+\Delta_{\alpha,1}^2r^{4-n}}}\mp\Lambda\right]^2
+4\left[\frac{\sqrt{c_1r+\Delta_{\alpha,1}^2r^{4-n}}}{r^2}\mp\Lambda\right]^2
} & \mbox{for}~\eqref{gs2}
\\
\displaystyle{ 8\left[\frac{c_1+\frac{16\alpha}{\kappa^2\mu^2}(1+\ln
r)} {4r\sqrt{c_1r+\frac{16\alpha}{\kappa^2\mu^2}r\ln
r}}\mp\Lambda\right]^2
+4\left[\frac{\sqrt{c_1r+\frac{16\alpha}{\kappa^2\mu^2}r\ln
r}}{r^2}\mp\Lambda\right]^2 } & \mbox{for}~\eqref{gs3}
\end{cases}
\, . \label{Kt}
\end{align}

As previously discussed for $n=4$, the above solution \eqref{gs2} is
unphysical for $n>3$ because the corresponding $\alpha$ becomes
negative. On the other hand, the solution of lower $n$ $(n=0,1,2)$
is obtained from positive $\alpha$. For \eqref{gs2} with $c_{1}\ge
0$, the solution is well defined in entire $r$ and no more
singularity except $r=0$. When $c_{1}<0$ and $n<3$, $B(r)$ in
\eqref{gs2} is real for $r\ge
r_{n}=\left(\frac{-c_{1}}{\Delta^{2}_{\alpha
,1}}\right)^{\frac{1}{3-n}}$ and the second line of \eqref{Kt} is
singular at $r=r_{n}$. When $c_{1}<0$ and $n>3$, $B(r)$ in
\eqref{gs2} is real for $r\le r_{n}$ and the second line of
\eqref{Kt} is singular at $r=r_{n}$. For \eqref{gs3} with positive
$\alpha$ and arbitrary $c_{1}$, $B(r)$ in \eqref{gs3} is real for
$r\ge r_{3}=\exp(\frac{-c_{1}\kappa^{2}\mu^{2}}{16\alpha})$ and the
fourth line of \eqref{Kt} becomes singular at $r=r_{3}$.

For \eqref{gs1}, there is no horizon for the upper plus sign, but the number
of horizons change from zero to two for the lower minus sign as
\begin{align}
\begin{cases}
\displaystyle{ r_{{\rm H}\pm}=-\frac{1}{2\Lambda}\left({\tilde
v}_{\lambda}\pm\sqrt{{\tilde v}_{\lambda}^2+4\Lambda}\right)},&
\quad\mbox{when}\quad \displaystyle{ {\tilde
v}_{\lambda}^2>-4\Lambda
}\\
r_{{\rm eH}}=\displaystyle{-\frac{{\tilde
v}_{\lambda}}{2\Lambda}}, & \quad\mbox{when}\quad
\displaystyle{{\tilde v}_{\lambda}^2= -4\Lambda}
\\
\mbox{no~horizon}, & \quad\mbox{when}\quad \displaystyle{{\tilde
v}_{\lambda}^2< -4\Lambda}
\end{cases}
\ ,
\end{align}
where
\begin{align}
{\tilde v}_{\lambda}=\sqrt{\frac{3\lambda-1}{3-\lambda}}\,\frac{4v}{\kappa\mu}.
\end{align}
In a similar fashion, it is easy to see from \eqref{gs2} that no
horizon exits for the upper plus sign, while horizons for the lower minus sign exist
and their structure is given as
\begin{align}
\begin{cases}
\displaystyle{ \mbox{two horizons}},& ~~\mbox{when}~~
\displaystyle{c_1>\frac{1}{r_{{\rm eH}}}-\left(2\Lambda+{\tilde
v}_{1}^2\right)r_{{\rm eH}} +\Lambda^2r_{{\rm eH}}^3
}\\
\displaystyle{ r_{{\rm eH}}=\sqrt{\frac{{\tilde
v}_{1}^2+2\Lambda+\sqrt{{\tilde v}_{1}^2+4{\tilde
v}_{1}^2\Lambda+16\Lambda^2}}{6\Lambda^2}}}\, , & ~~\mbox{when}~~
\displaystyle{c_1= \frac{1}{r_{{\rm eH}}}-\left(2\Lambda+{\tilde
v}_{1}^2\right)r_{{\rm eH}} +\Lambda^2r_{{\rm eH}}^3 }
\\
\mbox{\mbox{no~horizon}}, & ~~\mbox{when}~~ \displaystyle{c_1<
\frac{1}{r_{{\rm eH}}}-\left(2\Lambda+{\tilde v}_{1}^2\right)r_{{\rm
eH}} +\Lambda^2r_{{\rm eH}}^3 }
\end{cases}
\ ,
\end{align}

In the limit of Einstein gravity with the matter \eqref{M4}, we have
\begin{align}
B(r)=1-\frac{\Lambda}{2}r^2-\frac{M}{r}+\frac{4\alpha(1-3\lambda)}{(n-3)
\kappa^2\mu^2 \Lambda}r^{2-n},\qquad \delta(r)=\delta_0=0
\label{Ens}
\end{align}
which does not correspond to the solutions in HL gravity
\eqref{gs1}--\eqref{gs3} in the long distance limit. The matter
contribution in HL gravity goes as $r^{2-\frac{n}{2}}$ while that of
Einstein gravity follows $r^{2-n}$ as shown in \eqref{Ens}. This
noticeable difference arises form the nonlinear terms in the
Eqs. \eqref{Beq} and \eqref{deleq}, originated from higher order
curvature terms.

Let us now investigate field configurations that give the matter distribution
of power-law behavior \eqref{M4}.
We explore a few possible field configurations in detail and also discuss their
energy distributions mostly at long distance limit.

Because of electromagnetic duality of the Maxwell theory
\eqref{IRM}, the magnetic multipoles in the IR regime $(z=1)$ of the
HL U(1) gauge theory share the same pole structure with those of the
electric field.

At the Lifshitz UV fixed point, the theory has an anisotropic
scaling $z$ and magnetic potential in the U(1) theory takes the form which is
built out of  the detailed balance,
\begin{align}
-\frac{1}{4}F_{ij}F^{ij}\qquad
\stackrel{\mbox{IR}\rightarrow\mbox{UV}}{\Longrightarrow}\qquad
\begin{cases}
\displaystyle{-\frac{1}{8g_{z}^{2}}\left[(\nabla^{2})^{\frac{z-1}{2}}
F_{ij}\right]^{2}}~~\mbox{for~odd~}z\\
\displaystyle{-\frac{1}{8g_{z}^{2}}\left[(\nabla^{2})^{\frac{z}{2}}
\partial_{i}F_{ij}\right]^{2}}~~\mbox{for~even~}z
\end{cases}
.
\end{align}
The magnetostatic field of a magnetic monopole of charge $q_{\rm M}$
is obtained by solving the Bianchi identity with a $\delta$-function
singularity at the origin, $\epsilon_{ijk}\partial_i F_{jk}= q_{\rm
M}\delta^{(3)}({\vec{x})}$. By insertion of the monopole field
$\epsilon_{ijk} F_{jk}=q_{\rm M}x^i/4\pi r^3$ into the Lagrangian
density of U(1) gauge theory with anisotropic scaling $z$, we read
distribution of \eqref{M4} with $n=2(z+1)$. The results are
summarized in the second and third columns of Table 1. \vspace{4mm}

\begin{center}{
\begin{tabular}{|c|c|c|c|}
\hline\hline \multicolumn{1}{|c|}{$n$}& \multicolumn {3}{|c|}
{objects and scaling}\\
\cline{2-4} \multicolumn{1}{|c|}{(power)}& electric charge
&\multicolumn{1}{|c|}{magnetic monopole}
&\multicolumn{1}{|c|}{global monopole}\\
\cline{1-4}\hline\hline \multicolumn{1}{|c|}{$0$} &
\multicolumn{3}{|c|}{cosmological constant}
\\ \cline{1-4}
\hline\hline 2& & & $z=1$\\
4& monopole & $z=1$ & $z=2$\\
6&  & $z=2$ & $z=3$\\
8&  & $z=3$ & $z=4$\\
10& & $z=4$ & $z=5$\\[2mm]
\vdots & \vdots & \vdots & \vdots\\[2mm]
$n$&  & $\displaystyle{z=\frac{n}{2}-1}$
&$\displaystyle{z=\frac{n}{2}}$\\[1mm]
\hline\hline
\end{tabular}
}

\vspace{3mm} {\small Table 1. Electric, magnetic, and
global monopoles in field theories with various scaling $z$.}
\end{center}

\vspace{4mm}

We also consider other gravitating topological soliton,
global monopole, in the context of HL gravity. A scalar field
$\phi^{a}~(a=1,2,3)$ in the Lifshitz fixed point with an anisotropic
scaling $z$ \eqref{scaling} takes the following UV action in the
background metric \eqref{met1} with vanishing shift function
\begin{align}
S_{{\rm O(3)UV}}=\int dt\,d^{3}x\, \sqrt{g}N\,\left(
-\frac{1}{2N^{2}} \dot{\phi^{a 2}}-V_{{\rm UV}}\right),
\label{SUV}
\end{align}
where the potential in the UV regime is obtained by the detailed balance as
\begin{align}
V_{{\rm UV}}(\phi^{a},\partial_{i}\phi^{a},...)=
\begin{cases}
\displaystyle{
\frac{1}{8\kappa_{z}^{2}}
\left[\partial_{i}(\nabla^{2})^{\frac{z-1}{2}}\phi^{a}\right]^{2}}
~~\mbox{for~odd~}z\\
\displaystyle{
\frac{1}{8\kappa_{z}^{2}}
\left[(\nabla^{2})^{\frac{z}{2}}\phi^{a}\right]^{2}}
~~\mbox{for~even~}z
\end{cases}
.
\label{OUV}
\end{align}

Spacetime symmetry in HL gravity assumes to
be suddenly enhanced in the IR limit, i.e., the full diffeomorphism
symmetry is restored and the invariant light speed is recovered,
$x^{0}=ct=t$ in our unit system. The IR action is supposedly given by
\begin{align}
S_{{\rm O(3)IR}}=\int d^{4}x\sqrt{-g_{4}}\left(-\frac{g^{00}}{2}
\partial_{0}\phi^{a}\partial_{0}\phi^{a} -V_{{\rm IR}}\right),
\label{Oir}
\end{align}
where $g^{00}=1/N^{2}$ and the potential assumes to be of ordinary
quadratic spatial derivatives and of a quartic order self-interactions. The simplest potential that
gives rise to global monopoles is
\begin{align}
V_{{\rm IR}}(\phi^{a},\partial_{i}\phi^{a},...)
=-\frac{g^{ij}}{2}\partial_{i}\phi^{a}\partial_{j}\phi^{a}
-\frac{\lambda_{{\rm M}}}{4}(\phi^{2}-v^{2})^{2},\qquad
\phi^{2}\equiv \phi^{a}\phi^{a}.
\label{VIR}
\end{align}
This has a global O(3) symmetry, which is spontaneously broken to
O(2).

In the intermediate regime or along the flow line from the Lifshitz
UV fixed point to the relativistic IR fixed point, one expects that
kinetic term has the same form
\begin{align}
S_{{\rm O(3)}}=\int dt d^{3}x\sqrt{g}N\left(
-\frac{1}{2N^{2}}\dot{\phi}^{a2}-V\right),
\label{O3}
\end{align}
but the potential $V(\phi^{a},\partial_{i}\phi^{a},...)$ should be
connected to $V_{{\rm UV}}$ and $V_{{\rm IR}}$ at each UV and IR
limit. It is not clear whether the detailed valance can be attained
along with the flow.

We now attempt to find possible sources for \eqref{M4} from the O(3)
linear sigma model with various anisotropic scaling $z$. A constant
contribution, the $n=0$ case in \eqref{M4}, is easily identified as
either the symmetric vacuum $\phi=0$ in \eqref{VIR} or the core of
global monopole $r\le 1/(\sqrt{\lambda_{{\rm M}}}\,v)$ with
$\phi(r)\approx 0$, which yields
\begin{align}
\frac{\partial {\cal L}_{\rm M}}{\partial B}= 0, \qquad {\cal
L}_{\rm M}+\frac{\partial {\cal L}_{\rm M}}{\partial \delta}=
-\frac{\lambda_{{\rm M}}}{4}v^{4}.
\label{mat3}
\end{align}
This gives a positive contribution to the cosmological constant (see
also Table.~1), and then one is advised
to introduce the net
cosmological constant
\begin{align}
\Lambda_{v,\,\lambda}=\Lambda +\frac{2(1-3\lambda)\lambda_{{\rm M}}v^{4}}{3\Lambda \kappa^{2}\mu^{2}}.
\label{efco}
\end{align}
The corresponding solution in Einstein theory
\eqref{Beq2}--\eqref{deleq2} remains unchanged except
$\Lambda_{v,\,\lambda}$ instead of $\Lambda$. In HL gravity,
nontrivial solutions are
\begin{align}
B(r) = 1 -   \Lambda r^{2} \pm
\sqrt{\Lambda(\Lambda-\Lambda_{v,\,\lambda})}  \,r^{2} ,
\end{align}
which agrees with \eqref{gs1} for $n=0$.

It is worth pointing out that when the net cosmological constant $\Lambda_{v,\,\lambda}$ vanishes,
the vacuum solution of Einstein theory with vanishing integration constant reproduces flat space
as expected. On the other hand,  in HL gravity, we additionally obtain a nonconstant solutions
describing an AdS space due to nonlinearity
\begin{align}
B(r) = 1 -  2 \Lambda r^{2}.
\end{align}

To explore other possible configurations, we choose  a hedgehog ansatz
\begin{align}
\phi^{a}={\hat r}^{a}\phi(r)=(\sin\theta\cos\phi,\,
\sin\theta\sin\phi,\, \cos\theta) \phi(r),
\label{mnan}
\end{align}
to write
\begin{align}
S_{{\rm O(3)IR}}=4\pi\int_{-\infty}^{\infty}dt\int_{0}^{\infty}dr r^2
e^{\delta}\left[-\frac{B}{2}\phi'^{\,2}-\frac{\phi^2}{r^2}-\frac{\lambda_{{\rm
M}}}{4}(\phi^2-v^2)^2\right] ,
\end{align}
which yields
\begin{align}
\frac{\partial {\cal L}_{\rm M}}{\partial B}=&
-\frac{1}{2}\phi'^{\,2},
\label{mat1}\\
{\cal L}_{\rm M}+\frac{\partial {\cal L}_{\rm M}}{\partial \delta}=&
-\frac{B}{2}\phi'^{\,2}-\frac{\phi^2}{r^2}-\frac{\lambda_{{\rm
M}}}{4}(\phi^2-v^2)^2. \label{mat2}
\end{align}
We impose the boundary conditions of the scalar field as,
by requiring single-valuedness of the
field at the monopole position and finite energy at spacial infinity,
\begin{align}
\phi(0)=0,\qquad \phi(\infty)=v. \label{scbc}
\end{align}
A configuration satisfying both boundary conditions can be chosen such that
\begin{align}
\phi(r)=
\begin{cases}
0,&~ \mbox{for}~~ r\le \displaystyle \frac{1}{\sqrt{\lambda_{{\rm M}}}\,v},\\
v,&~ \mbox{for}~~ r>\displaystyle \frac{1}{\sqrt{\lambda_{{\rm
M}}}\,v},
\end{cases}
\end{align}
which is one that the scalar field has vacuum expectation value $v$ outside the monopole core
$r>1/(\sqrt{\lambda_{{\rm M}}}\,v)$. Near the vacuum, the
first term in \eqref{mat2} is negligible, and thus the leading term
approximation for the right-hand side of \eqref{mat1}--\eqref{mat2}
leads to
\begin{align}
\frac{\partial {\cal L}_{\rm M}}{\partial B}\approx  0, \qquad {\cal
L}_{\rm M}+\frac{\partial {\cal L}_{\rm M}}{\partial \delta}\approx
-\frac{v^{2}}{r^{2}},
\label{mat4}
\end{align}
which corresponds to the case $n=2$ and $\alpha=v^{2}$. Then the
spatial metric $B(r)$ for a gravitating global monopole is given by
either \eqref{gs1} or \eqref{gs2}.

We now discuss energy configurations near the Lifshitz fixed point.
If we suppose a global monopole of the ansatz \eqref{mnan}, satisfying the same boundary conditions
\eqref{scbc}, then we may read its energy in flat space and time
from \eqref{SUV}--\eqref{OUV}
\begin{align}
E\approx
\begin{cases}
\displaystyle{
\frac{\pi}{2\kappa_{z}^{2}}\int_{0}^{\infty}dr r^{2}
\left[\frac{d}{dr}\left(\frac{d^{2}}{dr^{2}}
+\frac{2}{r}\frac{d}{dr}
+\frac{1}{r^{2}}\right)^{\frac{z-1}{2}}\phi\right]^{2}}
~~\mbox{for~odd}~z
\\
\displaystyle{
\frac{\pi}{2\kappa_{z}^{2}}\int_{0}^{\infty}dr r^{2}
\left[\left(\frac{d^{2}}{dr^{2}}
+\frac{2}{r}\frac{d}{dr}
+\frac{1}{r^{2}}\right)^{\frac{z}{2}}\phi\right]^{2}}
~~~~~~~~\mbox{for~even}~z
\end{cases}
.
\end{align}
For large $r$, the leading term may also be $v^{2}/r^{2}$ term in
the parenthesis, and then the leading energy density behaves ${\cal
O}(1/r^{2z})$ at long distance limit. Comparing it to \eqref{M4}, we
have $n=2z$ and $\alpha\approx v^{2z}>0$, which seems true even in curved space and
time governed by the metric \eqref{rmet} (see also the fourth column
in Table.~1). An intriguing property of the $z>3/2$ global monopole
is possible finiteness of its energy in flat space and time because
Derrick's theorem is not applied due to the higher derivative terms
for anisotropic scaling.

To be consistent with Ho${\check{{\rm r}}}$ava-Lifshitz gravity with
$z=3$, we may take into account the scalar field theory with the
same scaling behavior, of which spatial derivative term may be sixth
order, $\partial_{i}\partial^{2}\phi\partial_{i}\partial^{2}\phi$.
The expected leading behavior of energy density of a global monopole
for large $r$ is of order of $1/r^{6}$, which is beyond the present
scope and should be dealt through further study.
Studying in detail global solitons is an attractive future research direction
of the field theories in the Lifshitz fixed point.

For an odd or arbitrary real positive $n$, we may need field
theories in the Lifshitz fixed point of fractional or arbitrary real
anisotropic scaling $z$. These field theories may not be easy to
deal with due to nonlocality, and so they may be realized in field theories at their UV fixed point.

\setcounter{equation}{0}
\section{Conclusion and Discussion}\label{sec5}

We have considered HL gravity with the detailed valance,
coupled with matters of power-law behaviors $1/r^{n}$, and shown
corresponding exact solutions. When both positivity of the matter energy density and
converging matter contributions at spatial infinity are required, the solutions exit only for a limited range of
$\lambda$.  Some cases, $n>4$, do not allow solutions for $\lambda=1$,
which is assumed to the limit to which $\lambda$ flows in the IR limit.
We also compared the long distance limit of the obtained metric $B(r)$ with
GR limit where the higher order terms in the action are neglected from
the beginning. We consistently found that if we keep the detailed balance,
then two limits do not coincide with each other, which may imply that the
detailed valance condition is to be modified to yield GR in the IR limit.

The most intriguing feature is that when HL gravity couples to the electric
field of a point charge, it naturally gives rise to a surplus solid angle,
which is hardly realized in GR unless exotic negative energy density is
introduced. As presented in Sec. 3, a surplus solid angle depends not
only on the parameters $\kappa\mu$ but also on $\lambda$. The valid range
of $\lambda$ for this case does not include $\lambda=1$. This indicates
that a surplus angle cannot emerge in the IR limit, but is likely formed
in the intermediate or UV regime where  $\kappa$ and $\mu$ can be
unconstrained.  Hence, this surplus angle is a unique feature of HL gravity
that may be used as a test of HL gravity.

We presented one of the possible effects of a surplus angle, sudden disappearance
(or appearance) of a star when an observer travels through a surplus area
in a folded geometry.
Considering the repulsive nature among the same charges, realization of
such object with accumulated charges seems unlikely. However, once it was
formed in early universe, it is likely inherited and its remnant might
be yet residing in the present universe.
Since the HL solution of surplus solid angle is not connected to the GR
solution of AdS charged black hole, we could not obtain a numerical value of the
surplus angle but an observation of surplus angle definitely supports
HL gravity as an astrophysical evidence.

It should be stressed that since there is no boot symmetry, a singularity seen in a given (static) frame does not guarantee  that other observers will see the singularity, especially for those who are in a relative motion to the static observer. This may alter the effects of surplus angles.  The sudden disappearance that we have described is based on   a source and observer in static configuration. When either one is moving, such effects should be reexamined with care, although surplus angle is a global property.

\section*{Appendix}\label{appendix}
\renewcommand{\theequation}{A-\arabic{equation}}
\setcounter{equation}{0}

We now analyze \eqref{Beq} and \eqref{deleq} in more detail.
Consider asymptotic behaviors of the solutions to
\eqref{Beq}--\eqref{deleq}. For sufficiently large $r$ at asymptotic
region, we assume the divergence of $B(r)$ arises as a power
behavior. Then some computation with the Eq. \eqref{deleq}
falls the leading behavior into two classes,
\begin{align}
B(r)\approx \left\{
\begin{array}{cll}
(\infty{\rm -i}) &-\Lambda r^{2} & \mbox{irrespective~of~~}\lambda \\
(\infty{\rm -ii}) & {\displaystyle B_{\infty}r^{p_{+}}} &
\mbox{for~~}\displaystyle{\lambda>1}
\end{array}
\right. , \label{ccso}
\end{align}
where the coefficient $B_{\infty}$ is an undetermined constant and
\begin{align}
p_{+}=\frac{2\lambda +\sqrt{2(3\lambda -1)}}{\lambda -1}\ .
\label{p+}
\end{align}
The long distance behavior in the first case ($\infty-$i) coincides
with the leading IR behavior in \eqref{deleq2} but the coefficient
is changed from $-\Lambda/2$ to $-\Lambda$. Since $p_{+}>2$ for
$\lambda>1$, the long distance behavior in the second case
($\infty-$ii) implies a possibility of new solution due to higher
order (and derivative) terms.

By the higher order terms for UV completion \eqref{LUV},
one expects that the short distance behavior is severely distorted
from $-M/r$ term in \eqref{deleq2}. The allowed powers for various
$\lambda$ are categorized as
\begin{align}
B(r)\approx \left\{
\begin{array}{cll}
(0{\rm -i}) & 1 & \mbox{irrespective~of~~}\lambda \\
(0{\rm -ii}) & c_{\frac{1}{2}} &
\mbox{for}~~\displaystyle{\lambda=\frac{1}{2}} \\
(0{\rm -iii}) & B_{0-}r^{p_{-}}~~\mbox{or}~~B_{0+} r^{p_{+}} &
\mbox{for}~~\displaystyle{\frac{1}{3}\le
\lambda<\frac{1}{2}} \\
(0{\rm -iv}) & B_{0+} r^{p_{+}} &
\mbox{for}~~\displaystyle{\frac{1}{2}<\lambda<1}
\end{array}
\right. , \label{ccsi}
\end{align}
where $B_{0\pm}$ are undetermined, $p_{+}$ is given in \eqref{ccso},
and $p_{-}$ is
\begin{align}
p_{-}=\frac{2\lambda -\sqrt{2(3\lambda -1)}}{\lambda -1}\ .
\label{p-}
\end{align}

The known exact static vacuum solutions \cite{Lu:2009em} can be
summarized and classified as how they connect two asymptotes with
various values of $\lambda$. The simplest solution independent of $\lambda$ is obtained by
adding unity and the cosmological constant term, $B(r)=1-\Lambda r^{2}$ and $\mbox{arbitrary}~~\delta(r),$
which connects
($\infty-$i) in \eqref{ccso} and ($0-$i) in \eqref{ccsi}.
For $\lambda\geq1/3$ and
$\lambda\ne 1$, one obtains other solutions, $B(r)=1-\Lambda r^{2}+B_{\pm} r^{p_{\pm}}$
and $\delta (r)= \frac{1+3\lambda \mp 2\sqrt{2(3\lambda-1)}}{1-\lambda}
\ln (r/r_{0})$, which connect one of
($\infty-$i)--($\infty-$ii) and one among ($0-$i) and
($0-$iii)--($0-$iv).
If we take the limit
$\lambda=1/3$ then $p_{+}=p_{-}=-1$, which is nothing but AdS
Schwarzschild solution with twice the cosmological constant. We note that the case
$\lambda=1/3$ in fact allows $B(r)=1-\Lambda r^{2}+\frac{M}{r}$ and arbitrary $\delta(r)$ since \eqref{Beq} vanishes in this limit.

A particular value of $\lambda$ also gives special solutions: For
$\lambda=\frac{1}{2}$ the solution is of the same form as the first solution independent of $\lambda$,
but allows any integration constant
$c_{\frac{1}{2}}$~\cite{Myung:2009dc}, $B(r) = -\Lambda \,r^{2} + c_{\frac{1}{2}}$ and $\delta(r) = \ln
(r/r_{0})$,
which connects ($\infty-$i) and ($0-$ii). For $\lambda=1$, one
obtains another solution connecting ($\infty-$i) and ($0-$i),
\begin{align}
B(r) = 1-\Lambda \,r^{2} \pm  \sqrt{c_{1} r} \ , \qquad
\delta(r)=\delta_{0}=0 \label{Bs3}
\end{align}
with an integration constant $c_{1}$. We can see that all the solutions above span possible boundary behaviors in \eqref{ccso} and \eqref{ccsi}.

\section*{Acknowledgments}
The authors would like to thank Chethan Krishnan, Jian Qiu,
and Bum-Hoon Lee for helpful comments.
The authors would like to thank Hang Bae Kim for valuable discussions.
This work was supported by the National Research Foundation of Korea(NRF)
grant funded by the Korea government(MEST) (No. 2009-0062869) through
Astrophysical Research Center for the Structure and Evolution of the Cosmos
(ARCSEC)) and
by the Korea Research Foundation Grant funded by the Korean Government
(KRF-2008-313-C00170) (Y.K.). S.K. is
supported by the Universit\'{e} Libre de Bruxelles and the
International Solvay Institutes, and by IISN - Belgium convention
4.4505.86 and by the Belgian Federal Science Policy Office through
the Interuniversity Attraction Pole P5/27.

\end{document}